\documentclass{emulateapj}

\usepackage{rotating}
\usepackage{lscape}

\usepackage{epsfig}
\usepackage{graphicx}

\shorttitle{Radio Observations of SBS\,0335$-$052}
\shortauthors{Johnson, Hunt, \& Reines}

\newcommand{\msun}{\ensuremath{~M_\odot}}

\newcommand{\lsun}{\ensuremath{~L_\odot}}

\newcommand{\hi}{H{\sc i}}
\newcommand{\hii}{H{\sc ii}}

\newcommand{\nuo}{{\ensuremath{\nu_0}}}

\newcommand{\alphant}{\ensuremath{\alpha_{\mathrm{nt}}}}
\newcommand{\tauff}{\ensuremath{\tau_{\mathrm{ff}}}}

\newcommand{\qlyc}{$Q_{Lyc}$}
\newcommand{\nesqv}{N$^2_{\rm e}\,V$}

\newcommand{\lbol}{$L_{\rm bol}$}

\newcommand{\bra}{${\rm Br}\alpha$}
\newcommand{\brg}{${\rm Br}\gamma$}
\newcommand{\pa}{${\rm Pa}\alpha$}
\newcommand{\ha}{${\rm H}\alpha$}

\newcommand{\lya}{${\rm Ly}\alpha$}

\newcommand{\sbs}{SBS\,0335$-$052}
\newcommand{\izw}{I\,Zw\,18}
\newcommand{\hen}{He\,2$-$10}
\newcommand{\hst}{{\it HST}}

\begin{document}

\title{Probing Star Formation at Low Metallicity: The Radio Emission of 
Super Star Clusters in SBS\,0335$-$052}

\author{Kelsey E. Johnson\altaffilmark{1}}
\affil{Department of Astronomy, University of Virginia, P.O. Box 3813,
    Charlottesville, VA 22904}
\email{kej7a@virginia.edu}

\author{Leslie K. Hunt}
\affil{INAF-Istituto di Radioastronomia-Sez. Firenze, L.go, Fermi 5, 
I-50125 Firenze, Italy}
\email{hunt@arcetri.astro.it}

\and

\author{Amy E. Reines}
\affil{Department of Astronomy, University of Virginia, P.O. Box 3813,
    Charlottesville, VA 22904}
\email{areines@virginia.edu}

\altaffiltext{1}{Adjunct at National Radio Astronomy Observatory, 520 Edgemont Road, Charlottesville, VA 22903, USA}

\begin{abstract}

  We present high-resolution radio continuum observations of the
  nascent starburst in the metal-poor galaxy \sbs.  These radio data
  were taken with the Very Large Array and include observations at
  0.7~cm, 1.3~cm, 2~cm, 3.6~cm, and 6~cm.  These observations enable
  us to probe the thermal radio nebulae associated with the extremely
  young star-forming regions in this galaxy.  Two discrete and
  luminous star-forming regions are detected in the south of the
  galaxy that appear to be associated with massive star clusters
  previously identified at optical wavelengths.  However, the
  remaining optically-identified massive star clusters are not clearly
  associated with radio emission (either thermal or non-thermal) down
  to the sensitivity limits of these radio data.  The spectral energy
  distributions of the two radio-detected clusters are consistent with
  being purely thermal, and the entire region has an inferred ionizing
  flux of $\sim 1.2 \times 10^{53}$s$^{-1}$, which is equivalent to
  $\sim 12,000$ ``typical'' O-type stars (type O7.5 V).  The
  observations presented here have resolved out a significant
  contribution from diffuse non-thermal emission detected previously,
  implying a previous episode of significant star formation.  The
  current star formation rate (SFR) for this southern region {\it
    alone} is $\sim 1.3$~M$_\odot$yr$^{-1}$, or $\sim
  23$~M$_\odot$yr$^{-1}$kpc$^{-2}$, which is nearing the maximum
  starburst intensity limit.  This SFR derived from thermal radio
  emission also suggests that previous optical recombination line
  studies are not detecting a significant fraction of the current star
  formation in \sbs.  From model fits to the radio spectral energy
  distribution, we infer a global mean density in the two youngest
  clusters of n$_e \gtrsim 10^3-10^4$~cm$^{-3}$.  In addition, a
  comparison between the compact and diffuse radio emission indicates
  that up to $\sim 50$\% of the ionizing flux could be leaking out of
  the compact \hii\ regions; this in is agreement with previous work
  which suggests that the interstellar medium surrounding the natal
  clusters in \sbs\ is porous and clumpy.

\end{abstract}
 

\section{Introduction}

\begin{figure*}[t!]
\epsscale{1}
\plotone{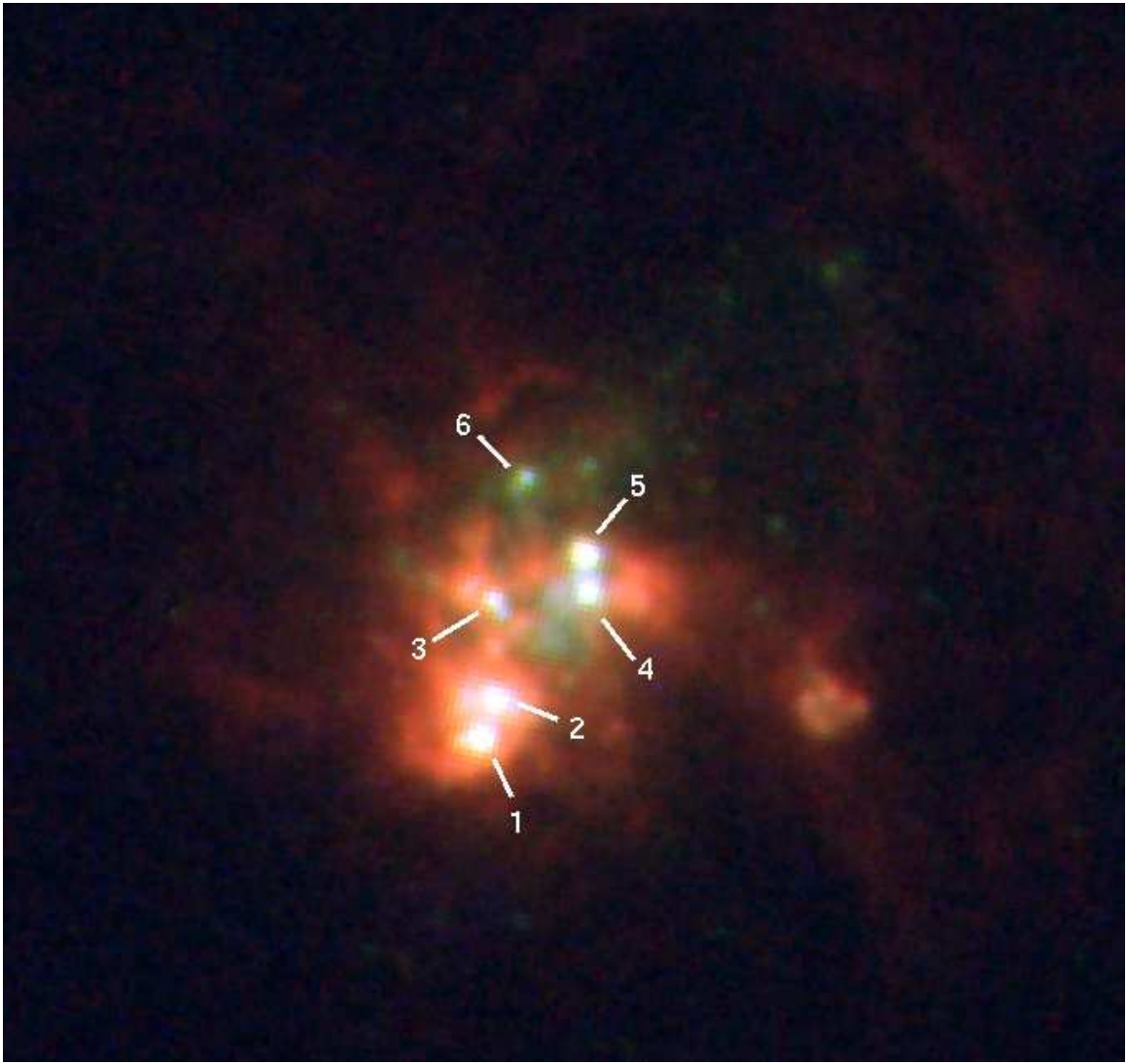}
\caption{Multicolor {\it Hubble Space Telescope} ACS image of
  \sbs\ constructed using the H$\alpha$ FR656N filter (red), the
  visual F550M filter (green), and the ultraviolet F220W$+$F330W
  filters (blue).  The SSCs identified by \citet{thuan97} are labeled.
  The image is shown in the standard orientation with North being the top
  of the image and is $\sim 10\arcsec \times 10\arcsec$.
}
\label{HST_image}
\end{figure*}
Super star clusters (SSCs) are the most massive and dense type of
young stellar clusters, and represent the most extreme mode of star
formation in the local universe.  SSCs are common in vigorous star
formation episodes \citep{hunter94,keto05}, apparently requiring high
pressure and interstellar gas column density for their formation
\citep[e.g.,][]{elmegreen97,billett02}.  A large amount of evidence
collected since the launch of the {\it Hubble Space Telescope}
suggests that at least some SSCs are the progenitors of globular
clusters; they have similar masses, sizes, and stellar density, and
evolutionary models suggest that evolved SSCs, if gravitationally
bound, would have properties very similar to present-day globulars
\citep{barth95,meurer95,ho96,larsen01,whitmore03}.  However, despite
these strong links between globular clusters and SSCs, we do not know
how the extremely low metallicity in the early universe affected
massive star cluster formation.  A low metal content may affect star
formation in a variety of ways, including cooling and pressure in the
birth cloud, hardness of the stellar spectra, and evolution of the
stars themselves
\citep[e.g.][]{schaerer02,smith02,tumlinson04,bate05}.  One way to
approach this question is to study SSC formation in extremely
low-metallicity systems in the local universe.

\sbs\ is a unique blue compact dwarf galaxy (BCD) in the nearby
universe because of its very high star formation rate (SFR) and its
extremely low metallicity; it is thus ideal for our study.  It was
discovered in the Second Byurakan Survey by \citet{izotov90} who noted
its particularly low oxygen abundance of $\sim 12 + {\rm log(O/H)} =
7.3$, confirmed by subsequent studies \citep{melnick92,izotov01}.
\sbs\ is the most metal-poor galaxy known with a SFR $\sim$1\,\msun
\,yr$^{-1}$, $\gtrsim$ 20 times higher than in
\izw\ \citep{fanelli88,hunt05b}.  \sbs\ is associated with an
\hi\ condensation within the same large \hi\ cloud
(64\,kpc$\times$24\,kpc) as \sbs W, which is located at a second
\hi\ peak about $\sim$22\,kpc to the west \citep{pustilnik01}.  The
neutral gas in \sbs\ is evidently not pristine, having roughly the
same oxygen abundance as the ionized gas \citep{thuan05}.  Analyses of
the broadband colors and spectral models of \sbs\ suggest that it may
be a local analog to primordial star forming sites in the early
universe \citep{thuan97,izotov97, papaderos98,pustilnik04}.

\begin{deluxetable*}{lcccccc}[t!]
\tabletypesize{\scriptsize}
\tablecaption{VLA Observations of \sbs\ \label{tab:obs}}
\tablewidth{0pt}
\tablehead{
\colhead{$\lambda$}&\colhead{Antenna}&\colhead{Date}&\colhead{Obs. Time }
&\colhead{Flux}& Program \\
\colhead{(cm)}&\colhead{Config.} &\colhead{Observed}&\colhead{(hours)}
&\colhead{Calibrator(s)} & Code}
\startdata
6   &  A-array      & 2003 Jun 03 & 2  & 3C48, 3C147 & AJ299 \\
6   &  A-array$+$PT & 2004 Oct 14 & 5  & 3C286   & AJ313\\
3.6 &  A-array      & 2003 Jun 03 & 2   & 3C48, 3C147 & AJ299 \\
3.6 &  A-array$+$PT & 2004 Oct 14 & 5  & 3C286   & AJ313\\
2   &  B-array      & 2003 Oct 29 & 7   & 3C48, 3C147 & AJ299\\
2   &  B-array      & 2004 Jan 02 & 5   & 3C48, 3C147 & AJ299\\
2   &  B-array      & 2004 Jan 07 & 4.5 & 3C48, 3C147 & AJ299\\
2   &  A-array      & 2004 Oct 11 & 4   & 0410+769 & AJ313\\
2   &  A-array      & 2004 Nov 30 & 3   & 0713+438 & AJ313 \\ 
2   &  A-array      & 2005 Jan 09 & 6.5 & 0410+769, 0713+438  & AJ313 \\
1.3 &  B-array      & 2004 Jan 04 & 6.5   & 3C48, 0410+769, 0713+438 & AJ299\\
1.3 &  B-array      & 2004 Jan 05 & 6   & 0410+769, 0713+438 & AJ299\\
1.3 &  B-array      & 2004 Jan 06 & 7   & 3C48, 0410+769, 0713+438 & AJ299\\
1.3 &  A-array      & 2004 Oct 12 & 7   & 3C286, 0410+769, 0713+438 & AJ313 \\
1.3 &  A-array      & 2004 Oct 16 & 7   & 3C286, 0410+769, 0713+438 & AJ313 \\
1.3 &  A-array      & 2004 Nov 30 & 3   & 0410+769, 0713+438 & AJ313 \\
0.7 &  B-array      & 2005 Mar 20 & 9   & 3C84, 3C286 & AJ313 \\
0.7 &  B-array      & 2005 Apr 03 & 7   & 3C84, 0410+769, 0319+415 & AJ313 \\
0.7 &  B-array & 2005 Apr 23 & 3.5 & 3C84, 3C286, 0410+769, 0713+438 & AJ313 \\
0.7 &  B-array & 2005 May 03 & 9   & 3C84, 3C286, 0410+769, 0713+438 & AJ313 \\
0.7 &  B-array & 2005 Jun 03 & 4.5 & 3C84, 3C286, 0410+769, 0713+438 & AJ313 \\
0.7 &  B-array & 2005 Jun 04 & 4.5 & 3C84, 3C286, 0410+769, 0713+438 & AJ313 \\
0.7 &  C-array      & 2005 Aug 11 & 4   & 3C84, 0410+769, 0713+438 & AJ313 \\
0.7 &  C-array      & 2005 Aug 12 & 9   & 3C84, 0410+769, 0713+438 & AJ313 \\
0.7 &  C-array      & 2005 Aug 13 & 9   & 3C84, 0410+769 & AJ313 \\
0.7 &  C-array      & 2005 Aug 14 & 4   & 3C84, 0410+769, 0713+438 & AJ313 \\
\enddata
\end{deluxetable*}

The star formation in \sbs\ is concentrated mainly in six
SSCs\footnote{Based on observations made with the Hubble
  Space Telescope, obtained from the Data Archive at the Space
  Telescope.  Science Institute, which is operated by the Association
  of Universities for Research in Astronomy, Inc., under NASA contract
  NAS 5-26555. These observations are associated with program
  \#10575.} (shown in Fig.~\ref{HST_image}), surrounded by a very blue
underlying envelope \citep{thuan97,papaderos98}.  Toward the north,
the filamentary and irregular structure of the emission suggests a
shell carved out by supernovae (SNe).  The SSCs (1-6, according to the
notation of \citealt{thuan97}) are roughly aligned in the southeast to
northwest direction.  \citet{reines08} confirm an age gradient of the
clusters across the galaxy (with the northern-most clusters being the
oldest) that is consistent with a large-scale disturbance crossing the
galaxy at a speed of $\sim 35$~km~s$^{-1}$ and triggering star
formation en route.

Although the optical colors of the two youngest SSCs 1$+$2 are
relatively blue, they are also strong infrared (IR) emitters
\citep{thuan99,hunt01,dale01,plante02,houck04,reines08}.  Extinction
estimates around these two clusters vary greatly depending on the
wavelengths of the observations used.  Low extinctions (A$_V \sim
0.5$\,mag) are derived from optical observations
\citep[e.g.,][]{izotov97,reines08} whereas mid-IR observations
indicate high extinction (A$_V\,\gtrsim\,$12\,mag) regions surrounding
the young clusters \citep[e.g.,][]{thuan99, plante02,hunt05a,
  houck04}.  \citet{reines08} find evidence for a porous and clumpy
ISM surrounding the young clusters, which can naturally account for
the apparently discrepant extinction estimates found in the
literature.  Dense dust clumps heated by the impinging UV stellar
continuum provide the high extinction estimates from mid-IR
observations, whereas diffuse inter-clump regions sampled by optical
observations yield low measured extinctions.

Previous low-resolution radio observations of \sbs\ showed free-free
absorption on a global scale, and a significant non-thermal component
\citep{hunt04}.  However, the beam size of those images was
insufficient to disentangle the spatial distribution of the thermal
and non-thermal emission.  In this paper, the radio continuum
properties of \sbs\ are re-examined at significantly higher resolution
with the goal of probing individual star-forming regions.  We present
new high-resolution radio continuum observations of \sbs\ to better
separate the different emission mechanisms spatially, and probe the
thermal radio nebulae.  We describe our observations in
in $\S$2, discuss the morphology of the natal
clusters in $\S$3, and analyze the physical
properties of the clusters in $\S$4. The relationship
between the clusters and the overall galaxy are presented in
$\S$5, and $\S$6 provides a general
summary of the results.  We adopt a distance of 55.7\,Mpc to \sbs,
which assumes $H_0\,=\,70$\,km\,s$^{-1}$\,Mpc$^{-1}$, and correction
to the CMB reference frame as described in \citet{hunt05a}.  This
corresponds to a spatial scale of 270~pc~arcsec$^{-1}$.

\section{Observations \label{sec:observations}}

High-resolution radio observations of \sbs\ were obtained with
the Very Large Array (VLA)\footnote{The National Radio Astronomy
Observatory is a facility of the National Science Foundation operated
under cooperative agreement by Associated Universities, Inc.} from
2003 June to 2005 August.  Observations were obtained at C-band (5
GHz, 6 cm), X-band (8 GHz, 3.6 cm), U-band (15 GHz, 2 cm), K-band (22
GHz, 1.3 cm), and Q-band (43 GHz, 0.7 cm); these are summarized in
Table~\ref{tab:obs}.  The high-frequency observations at K-band and Q-band
utilized fast-switching to a nearby phase calibrator with cycle times
of $\sim 2$ minutes in order to mitigate the effect of atmospheric
changes.

These radio data were reduced and calibrated using the Astronomical
Image Processing System (AIPS).  When available, models of the flux
calibrators were used in order to exploit the full $uv$ coverage of
these sources.  Most of the data sets included observations of more
than one flux calibrator in order to cross check the calibration
solutions.  We estimate that the absolute uncertainty in flux
calibration is $\lesssim 5$\%, based on scatter in the VLA Flux Density
Calibrator database for the calibrators used in this program.

Because our aims depend on relative flux densities at different
frequencies, care was taken to obtain the best-matched $uv$ coverage
possible at each frequency.  Each frequency was observed in two
separate array configurations chosen to obtain relatively well-matched
synthesized beams. Additionally, all of the combined data sets were
trimmed to have an identical minimum $uv$ spacing of 30~k$\lambda$ in
order to roughly match their sensitivity to extended structure, and
the ``robust'' parameter was varied between -1 and 1 (slightly uniform
to slightly natural) in order to obtain better matched synthesized
beams.  Nevertheless, the $uv$ coverage at each frequency cannot be
perfectly matched, and thus there may be slight variations between the
sensitivity to different spatial scales at each frequency.  An
additional set of images was also made using the greatest possible
sensitivity at each frequency using purely natural weighting
(robust=5) and no restrictions in $uv$ coverage.  The resulting
imaging parameters are listed in Table~\ref{tab:imaging}.

Unfortunately, the highest frequency observations at Q-band were
consistently plagued by bad weather and instrumental problems.  As a
result the rms noise in the combined Q-band data set is too high to be
useful for the purposes of this project, and is not analyzed further
in this paper.  However, in the final data sets obtained
in August 2005, a luminous transient object appeared, which will be
discussed in a separate paper.

Determining the flux densities of non-point-like sources observed with
an interferometer amid complex backgrounds is notoriously prone to
large uncertainties.  In order to obtain the most accurate values
possible, we employed several techniques to measure the flux
densities.  First, the peak flux densities were measured in mJy/beam;
if the objects are point-like, this is equal to their total flux
density in mJy.  Second, aperture photometry was performed on each
wavelength using a variety of identical (irregular) apertures and
annuli using the AIPS++ viewer tools.  Finally, the sources were fit
with Gaussian profiles using the AIPS task JMFIT.  The quoted flux
densities and uncertainties in Table~\ref{tab:fluxes} reflect these
results.  In addition, photometry was performed on the high-sensitivity
images using a $0.5''$ radius aperture ($=135$~pc) centered on the
entire region of radio emission ``region S'', and these results are
also quoted in Table~\ref{tab:fluxes}.

\begin{deluxetable}{ccccc}
\tabletypesize{\scriptsize}
\tablecaption{Imaging Parameters \label{tab:imaging}}
\tablewidth{0pt}
\tablehead{
\colhead{$\lambda$} &\colhead{Weighting} &\colhead{Synth. Beam} &\colhead{P.A.}
 &\colhead{rms}\\
\colhead{(cm)}&\colhead{robust value}&\colhead{($''\times''$)} & 
\colhead{($\degr$)} &\colhead{($\mu$Jy/beam)}}
\startdata
6 &   -1  & $0.38\times0.19$ & 25 & 18 \\
6 &    5  & $0.53\times0.40$ & -1 & 10 \\

3.6 & 0   & $0.23\times0.18$ & 9 & 12 \\
3.6 & 5   & $0.31\times0.26$ & -14 & 10 \\

2.0 & 0.5 & $0.25\times0.20$ & 17 & 17 \\
2.0 & 5 & $0.39\times0.34$ & 6  & 15 \\

1.3 & 1   & $0.24\times0.20$ & 3 & 12 \\
1.3 & 5   & $0.29\times0.24$ & 2 & 11 \\

0.7 & 5   & $0.23\times0.18$ & 34 & 40 \\

\enddata
\end{deluxetable}

\begin{deluxetable}{ccccc}
\tabletypesize{\scriptsize}
\tablecaption{Radio Flux Densities  \label{tab:fluxes}}
\tablewidth{0pt}
\tablehead{
&\colhead{$F_{1.3\rm{cm}}$} &\colhead{$F_{2.0\rm{cm}}$} 
&\colhead{$F_{3.6\rm{cm}}$} &\colhead{$F_{6.0\rm{cm}}$}\\
\colhead{Source}&\colhead{($\times 10^{-4}$ Jy)}&\colhead{($\times 10^{-4}$ Jy)} & 
\colhead{($\times 10^{-4}$ Jy)} &\colhead{($\times 10^{-4}$ Jy)}}
\startdata
S1 & $1.7\pm0.3$ & $1.8\pm0.5$ & $1.8\pm0.4$ & $1.4\pm0.2$  \\
S2 & $1.2\pm0.2$ & $1.0\pm0.2$ & $0.8\pm0.2$ & $0.8\pm0.2$  \\
Region S\tablenotemark{a} & $5.6\pm0.6$ & $5.8\pm0.6$ & $4.6\pm0.5$ 
& $3.4\pm0.4$\\
\enddata
\tablenotetext{a}{Region S corresponds to an aperture of $0.5''$ radius 
centered on the radio emitting region that encompasses S1 and S2. }
\end{deluxetable}

\section{Location and Morphology of the Radio Emission \label{sec:morphology}}

\begin{figure*}[t!]
\hbox{
\includegraphics[angle=0,width=0.5\linewidth,bb=60 160 550 610]{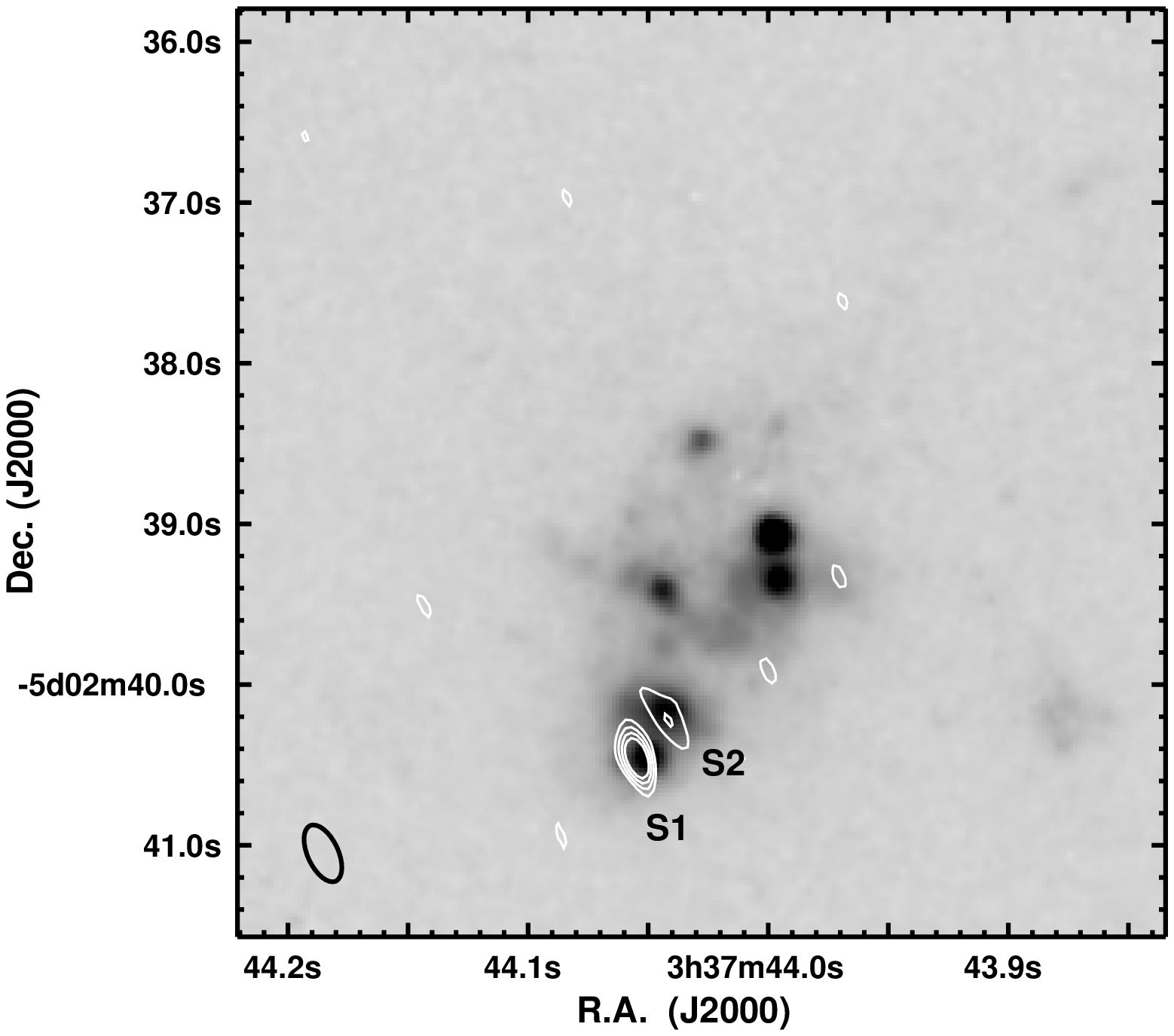}
\includegraphics[angle=0,width=0.5\linewidth,bb=60 160 550 610]{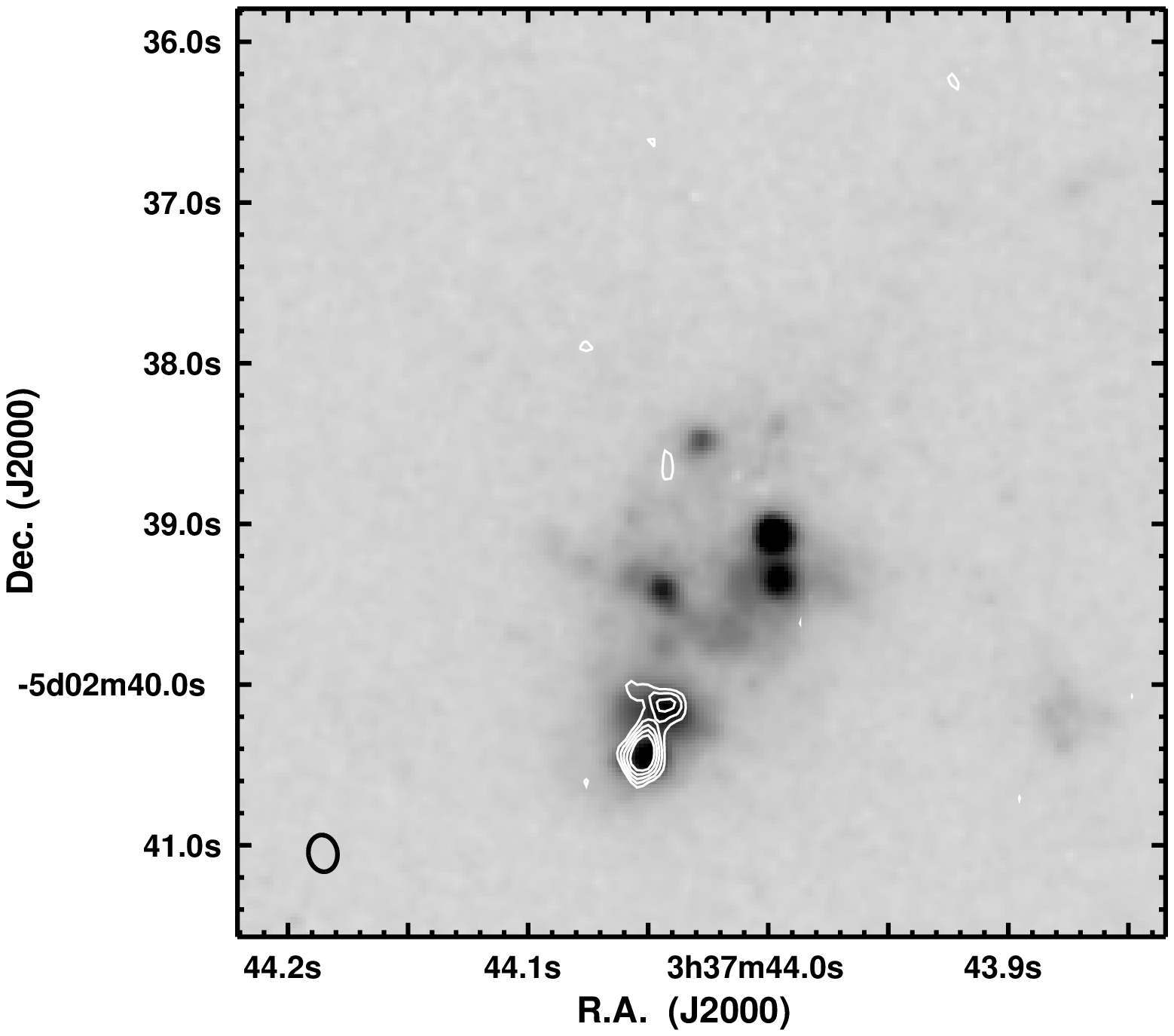}
}
\hbox{
\includegraphics[angle=0,width=0.5\linewidth,bb=60 160 550 650]{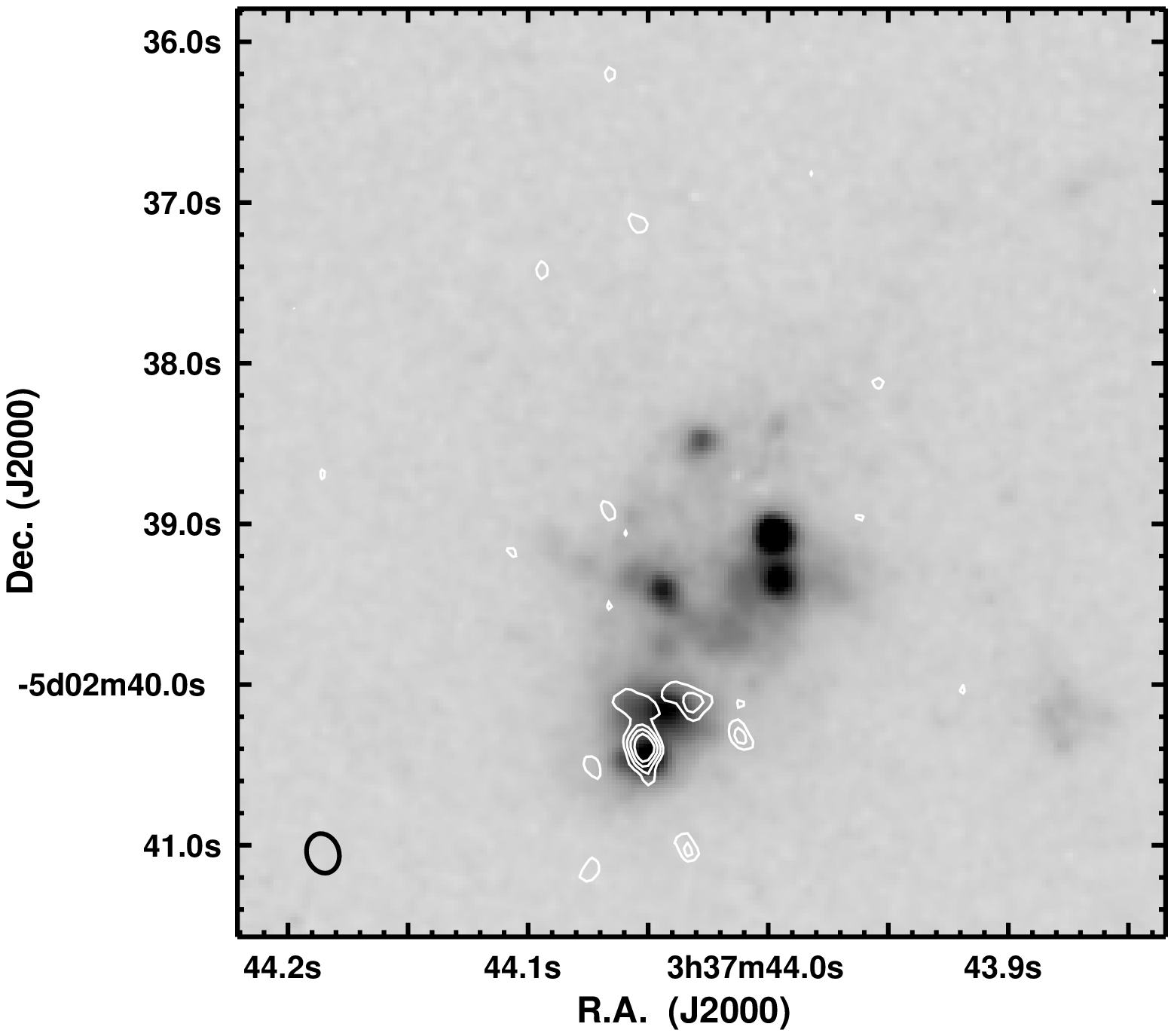}
\includegraphics[angle=0,width=0.5\linewidth,bb=60 160 550 650]{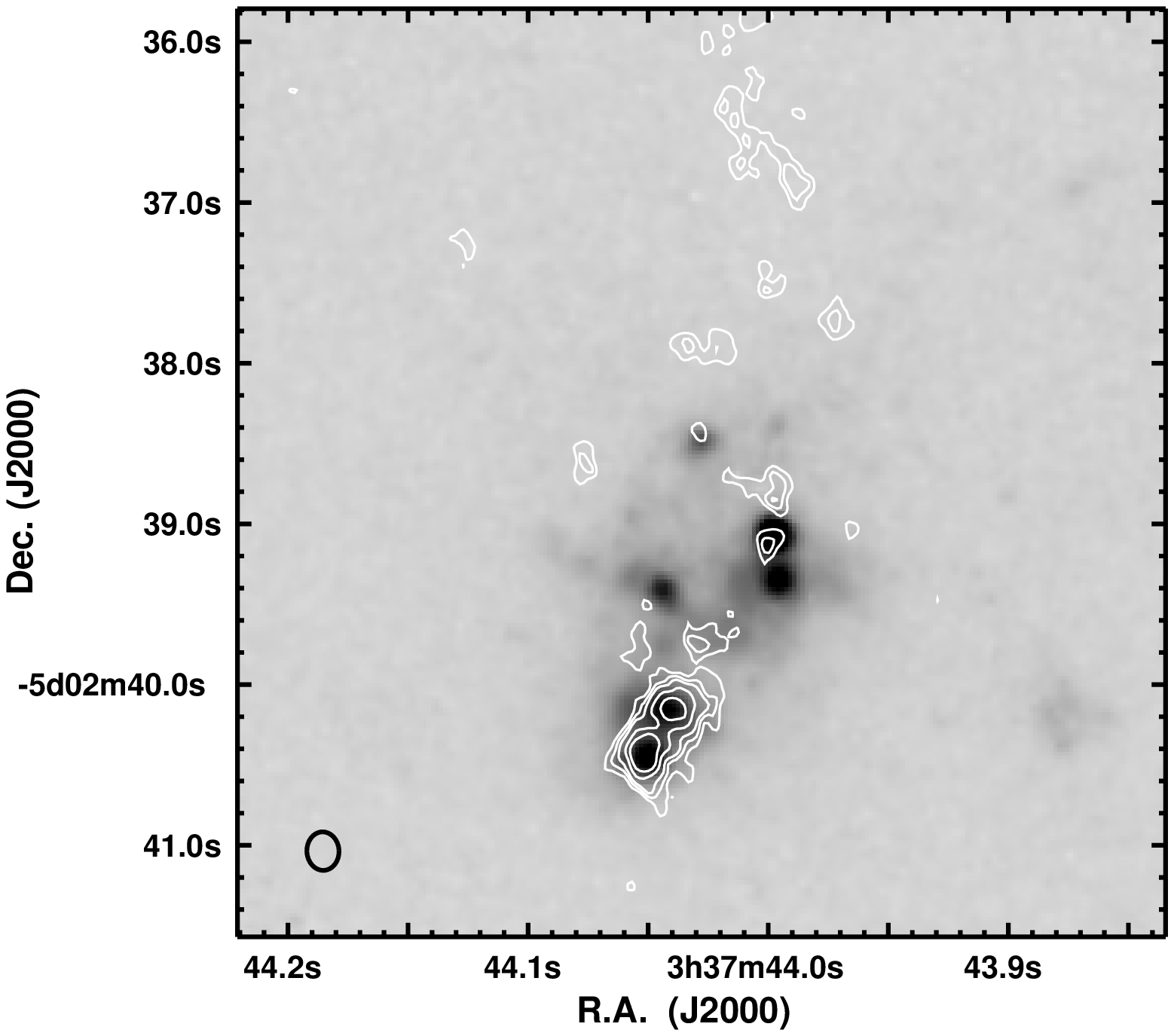}
}
\caption{VLA contours overlaid on an HST ACS F550M image in gray-scale.
{\bf (a)}: VLA 6~cm contour levels of $3, 4, 5, 6 \times \sigma$ (18~$\mu$Jy/beam). 
{\bf (b)}: VLA 3.6~cm contour levels of $3, 4, 5, 6, 7 \times \sigma$ (12~$\mu$Jy/beam).
{\bf (c)}: VLA 2~cm contour levels of $3, 4, 5, 6 \times \sigma$ (17~$\mu$Jy/beam). 
{\bf (d)}: VLA 1.3~cm contour levels of $3, 4, 5, 7, 9 \times \sigma$ (11~$\mu$Jy/beam).  \label{fig:XCUKcontV}}
\end{figure*}

The compact radio emission in \sbs\ is resolved into two dominant
sources S1 and S2 located in the southern region of the galaxy that
appear to be associated with two SSCs 1 and 2 identified by
\citet{thuan97}.  The radio emission at 1.3~cm, 2~cm, 3.6~cm, and 6~cm
is shown overlaid on an \hst/ACS F550M image in
Figure~\ref{fig:XCUKcontV}.

The exact location of the radio sources relative to the optical SSCs
is difficult to pinpoint precisely because of the uncertainty in the
\hst\ astrometry.  The astrometry for the VLA images is accurate to
within $\sim$~0\farcs1 and is considered absolute by comparison.  We
started first with the same \hst\ image as in \citet{hunt04},
astrometrically calibrated with five stars in the USNOA2.0.  The
astrometric rms uncertainty is 0\farcs54, roughly twice the size of
the highest resolution synthesized beams of the radio data presented
here.  Using the nominal astrometry, the southernmost radio peak lies
$\sim$0\farcs4 south of the two brightest SSCs.

We then attempted to improve the relative pointing uncertainty between
the optical and radio images by bootstrapping from the morphology of
the Pa$\alpha$ emission presented in \citet{reines08}.  Using this
method, we first matched the continuum-subtracted Pa$\alpha$ emission
to the radio, and applied that astrometric solution to the
non-continuum-subtracted Pa$\alpha$ image, which has many features in
common with the optical image.  Subsequently, the optical image was
aligned with the non-continuum subtracted Pa$\alpha$ image to obtain
the final astrometry.  
The shift in the astrometry relative to the USNO solution is $\sim$0\farcs4,
within the expected uncertainty of the absolute positions. 
The resulting alignment of the optical image
clearly associates the radio sources S1 and S2 with the optically
identified SSCs 1 and 2 as shown in Figure~\ref{fig:XCUKcontV}.  Given
this apparent correspondence between the optical and radio sources,
hereafter we refer to S1 and S2 as SSCs 1 and 2.

 \begin{figure*}[t!]
\epsscale{1.}
\plottwo{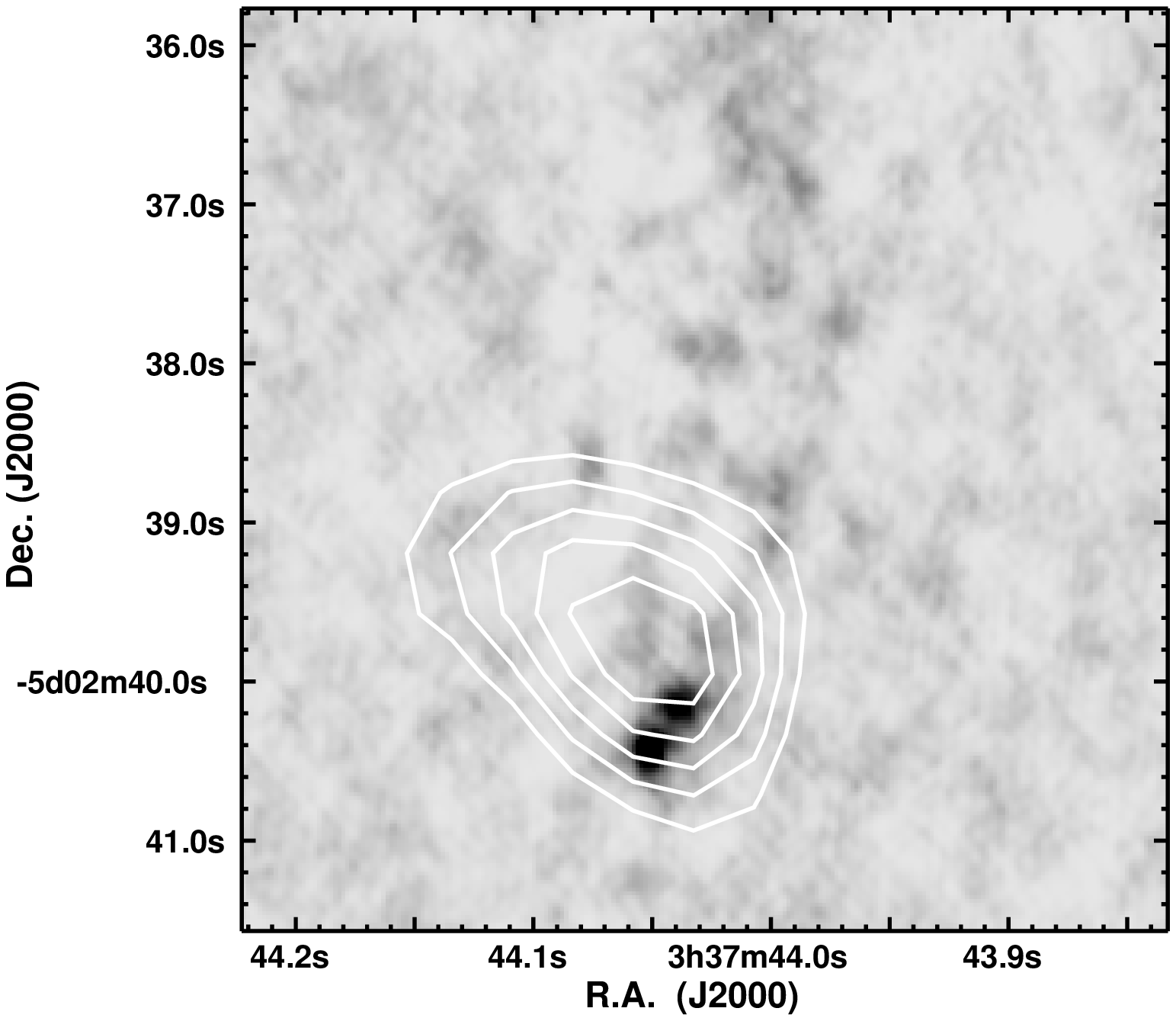}{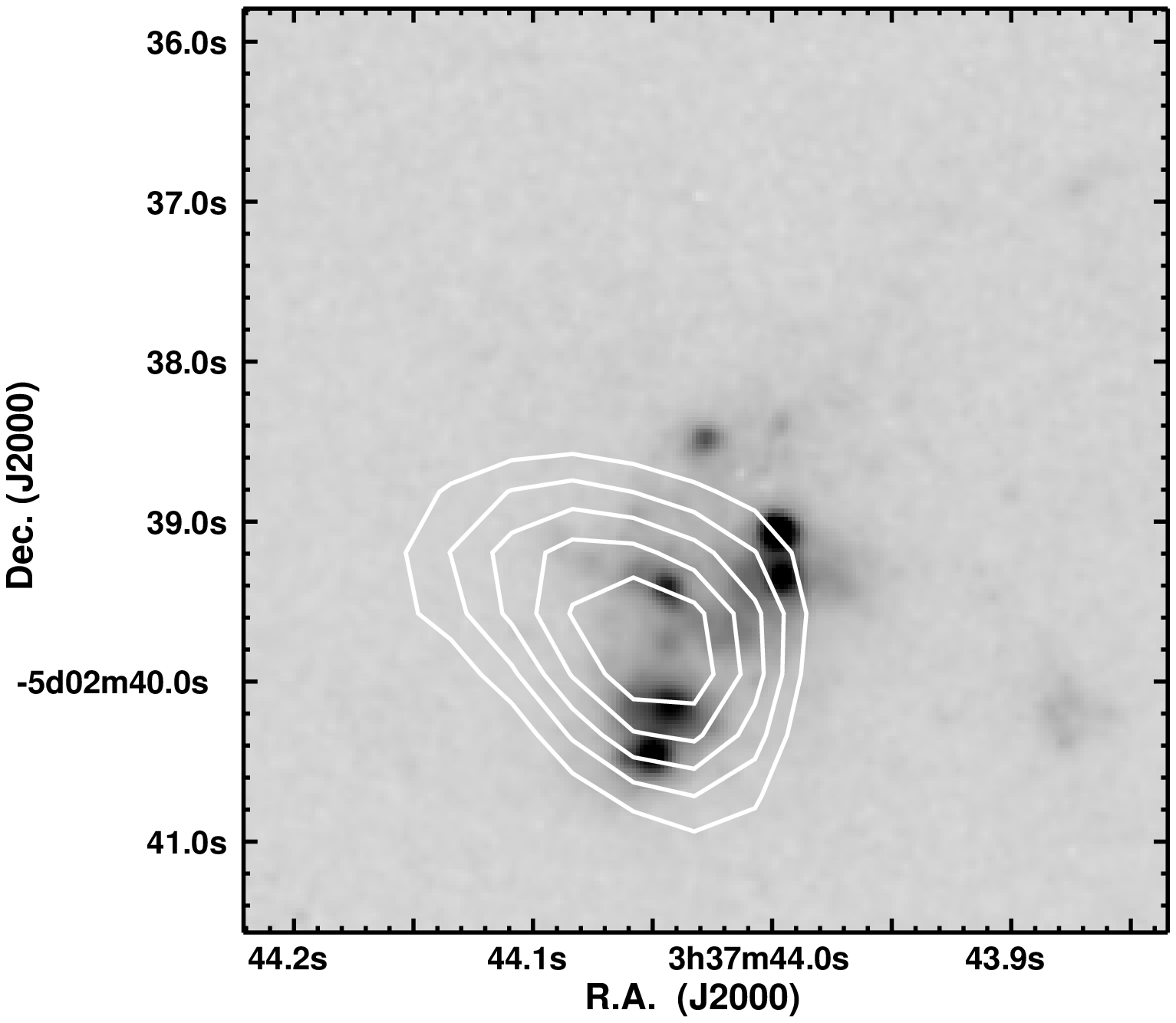}
\caption{L-band (20~cm) contours from \citet{hunt04} overlaid on
  (left) the K-band (1.3~cm) image of SBS~0335-052 and (right) the ACS
  F550M image of SBS~0335-052.  Although at lower resolution, the
  centroid of the 20~cm source is offset from the thermal emission by
  $\sim$~0\farcs5 to the north.
\label{LcontK}}
\end{figure*}

None of the other SSCs identified optically by Thuan et al. have
firmly-detected radio emission, although SSCs 5 and 6 have possible
marginal detections in the 1.3\,cm K-band image.  In addition, there
is a string of K-band $\sim 3-5 \sigma$ peaks extended toward the
north of \sbs\ (see Fig. \ref{fig:XCUKcontV}).  Individually, any one
of these peaks would not necessarily be considered a source; however,
the spatial correlation is suggestive of a more extended feature that
is largely resolved out by these high-resolution observations.  We
hypothesize that this feature may be associated with the supershell
observed by \citet{thuan97} and/or an outflow extending in that
direction.

The flux density at high frequencies observed for Region~S in this
paper compared to the previous work at lower resolution \citep{hunt04}
suggests that even in these high-resolution observations the spatial
filtering inherent in array observations is not ``resolving out'' any
of the thermal flux.  At 1.3\,cm and 2\,cm, our high-resolution observations
recover all the flux in \sbs\ reported by \citet{hunt04}.  This
implies that the size of the entire thermal source at high frequencies
must be smaller than or comparable to the 0.5\arcsec aperture used to measure
the flux density of Region~S.  

In contrast, the 20~cm radio emission at a resolution of 1\farcs6
presented by \citet{hunt04} is peaked $\sim$0\farcs5 ($\sim$100\,pc)
north of sources SSCs 1 and 2.  Fig.~\ref{LcontK} illustrates the
overlay of the 20~cm emission on the 1.3~cm map\footnote{The
  positional discrepancy between this work and \citet{hunt04} is a
  consequence of the new \hst\ images and astrometry presented in this
  paper.}.  Because the low-frequency radio flux is dominated by
synchrotron emission, this non-thermal emission is likely to be
associated with SNe from a distinct episode of star formation with an
age of $\gtrsim$3.5\,Myr.  At 6\,cm, the lowest frequency of the
observations presented here, we recover only 44\% of the global flux
measured by \citet{hunt04}.  This would imply that the non-thermal
emission is much more diffuse than the thermal; indeed if the
significant synchrotron halo of I\,Zw\,18 \citep{hunt05b}, another
low-metallicity BCD, were placed at the distance of \sbs, it would be
resolved out by our observations (see $\S$\ref{sec:nt}).

\section{Physical Properties of the Natal SSCs \label{sec:fits}}

\subsection{Modeling the Spectral Energy Distributions}

To gain insight into the physical conditions in the radio sources
present in \sbs\, we apply a general thermal$+$non-thermal fit with an
absorption component as in \citet{hunt04}.  These are simple models
that assume a uniform-density homogeneous ionized medium.  The
realistic physical conditions in the \hii\ regions are likely to be
more complex, including density variations and deviations from a
sphere, but the available data do not support more complicated models.
We can express the thermal radio free-free absorption coefficient
$\kappa_\nu$ as:
\begin{equation}
\kappa_\nu\,\simeq\,0.08235\left(\frac{T}{\rm K}\right)^{-1.35}
\left(\frac{n_e}{\rm cm^{-3}}\right)^2
\left(\frac{\nu}{\rm GHz}\right)^{-2.1}\quad {\rm pc^{-1}}
\end{equation}
where 
$n_e$ is the electron density of the ionized gas,
$T$ is the ionized gas temperature, and
$\nu$ is the frequency. 
This means that the free-free optical depth \tauff\,=\,$\kappa_\nu \,L$ 
($L$$\sim$path length through the region) is:
\begin{equation}
\tau_{\rm ff}\,\simeq\,0.08235\left(\frac{T}{\rm K}\right)^{-1.35}
\left(\frac{\rm EM}{\rm pc\,cm^{-6}}\right)
\left(\frac{\nu}{\rm GHz}\right)^{-2.1}	
\end{equation}
where EM is the linear emission measure ($\sim n_e^2\,L$). 

By virtue of Kirchhoff's law for thermal emission, which
relates $\kappa_\nu$ and the radio volume emissivity $j_\nu$, 
we write the thermal radio flux as a function of EM:

\begin{eqnarray}
f^{\rm thin}_\nu\ & =\  \phi\ (5.95\times10^{-5})\
\left(\frac{T}{\rm K}\right)^{-0.35} 
\left(\frac{\nu}{\rm GHz}\right)^{-0.1} \nonumber \\
& \times \left(\frac{EM}{\rm pc\ cm^{-6}}\right)
\left(\frac{\theta}{\rm arcsec}\right)^2 \quad {\rm mJy}
\end{eqnarray}


$\phi$ is dimensionless and depends on the geometry, equal to $\pi/6$
for a spherical region of constant density and diameter (FWHM)
$\theta$, and $\pi/4$ for a cylindrical region of diameter and length
$\theta$ (e.g., \citealt{mezger67}).  $\phi$ also accommodates a
filling factor, which here is assumed equal to unity.  We take into
account the optical thickness of the radio emission through
\tauff\ assuming a geometry in which the emission and absorption are
intermixed in a homogeneous medium.  The thermal radio flux
$f^{th}_\nu$ can then be expressed as:

\begin{eqnarray}
&f^{th}_\nu  =  \left[\frac{1 - \exp(-\tauff)}{\tauff}\right]\ f^{\rm thin}_\nu  \nonumber &\\
 =  &\left[1 - \exp(-\tauff)\right]
\phi\ (7.225\times10^{-4})\
\left(\frac{T}{\rm K}\right)
\left(\frac{\nu}{\rm GHz}\right)^{2} \nonumber \\
 &\times \left(\frac{\theta}{\rm arcsec}\right)^2 \quad {\rm mJy}
\label{eqn:thermal}
\end{eqnarray} 

Equation \ref{eqn:thermal} shows that in the limit of
optically-thin radio emission where the absorption term $\left[1 -
  \exp(-\tauff)\right] \propto$ \tauff, $f^{th} \propto \nu^{-0.1}$.
Alternatively, when \tauff$\,\gg 1$, the radio emission is completely
optically thick, and $f^{th} \propto \nu^{2}$.  Hence we are not
making any assumptions a priori about the nature of the thermal radio
emission\footnote{These are the same models as in \citet{hunt04}, but
  expressed in a more general way.}.  A spherical geometry is assumed
for the thermal emission, and the electron temperature $T_e$ was taken
to be 20000\,K \citep{izotov99}.  

Allowing for the most general models that also include non-thermal
emission introduces two additional free parameters (the non-thermal
flux density and spectral slope), leaving the models unconstrained.
Nevertheless, for completeness we also modeled these regions with the
inclusion of a non-thermal component to assess the degree to which the
results might be affected.  We assume a screen geometry for the
free-free absorption and write for the non-thermal radio flux
$f^{nt}_\nu$:
\begin{equation}
f^{nt}_\nu\ =\ \exp(-\tauff) 
\ f^{\rm nt}_{\nuo}\left(\frac{\nu}{\nuo}\right)^{\alphant} \quad {\rm mJy}
\label{eqn:nonthermal}
\end{equation} 
where $f^{\rm nt}_{\nuo}$ is the non-thermal (unabsorbed) flux at
frequency \nuo.  The total flux in this case at given frequency $\nu$
is the sum of Equations \ref{eqn:thermal} and \ref{eqn:nonthermal}.

The data were fit to models allowing for both thermal and non-thermal
contributions using a $\chi^2$ minimization algorithm, and contour
values of the reduced $\chi^2$ values and resulting spectral energy
distributions (SEDs) are shown in Figures~\ref{plot_S1}a-c.  As in
\citet{hunt04}, we fit the data using three free parameters: $f^{\rm
  nt}_{\nuo}$, diameter $\theta$, and emission measure EM.  We
ran models separately for values of \alphant\ ranging from $-0.45$ to
$-0.8$, which would enable us to judge, albeit crudely, the best-fit
non-thermal index from the lowest $\chi^2$.  The best-fit parameters
derived from these models are listed in Table~\ref{tab:properties} and
discussed in the following.  In each case, the radio flux densities
are consistent with a purely thermal origin.

\subsection{Radii and Densities \label{sec:fits}}
\begin{figure*}[t!]
\epsscale{0.7}
\plotone{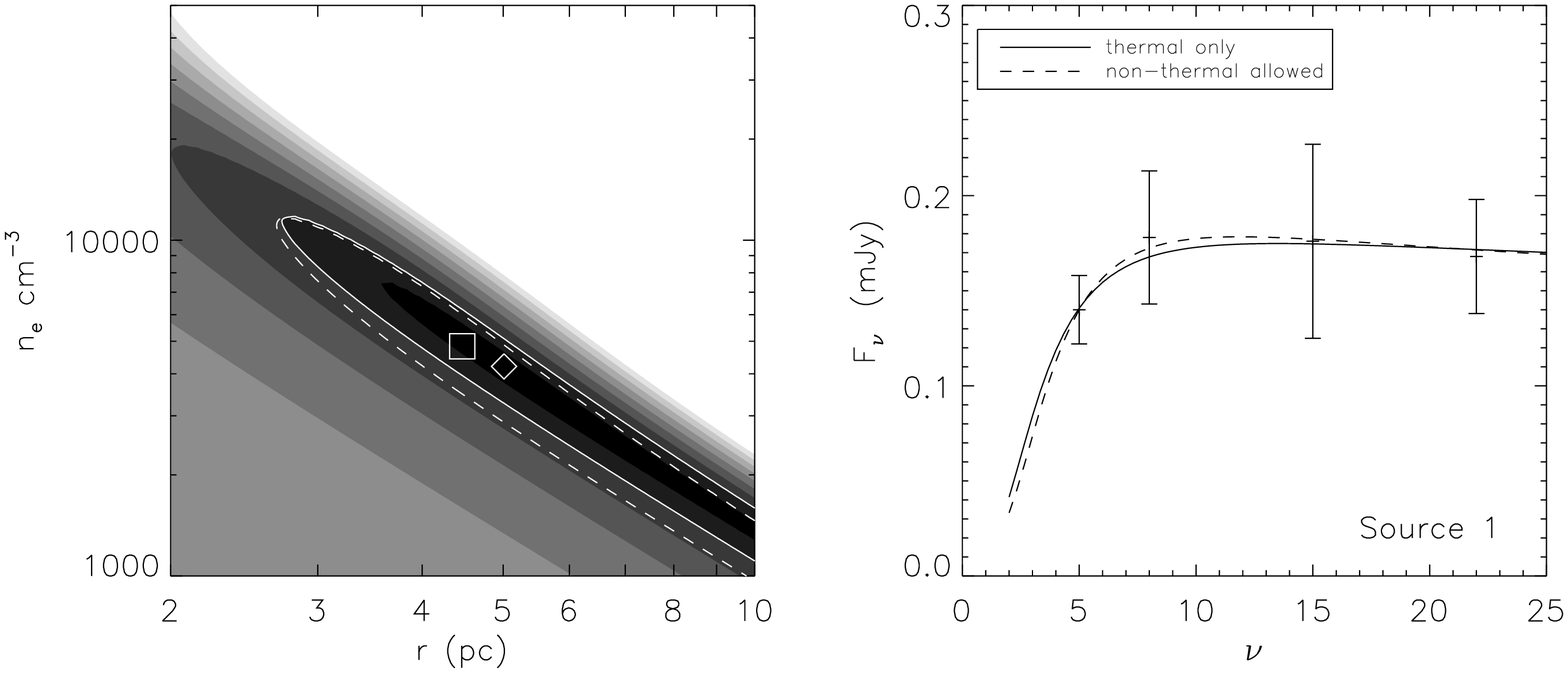}
\plotone{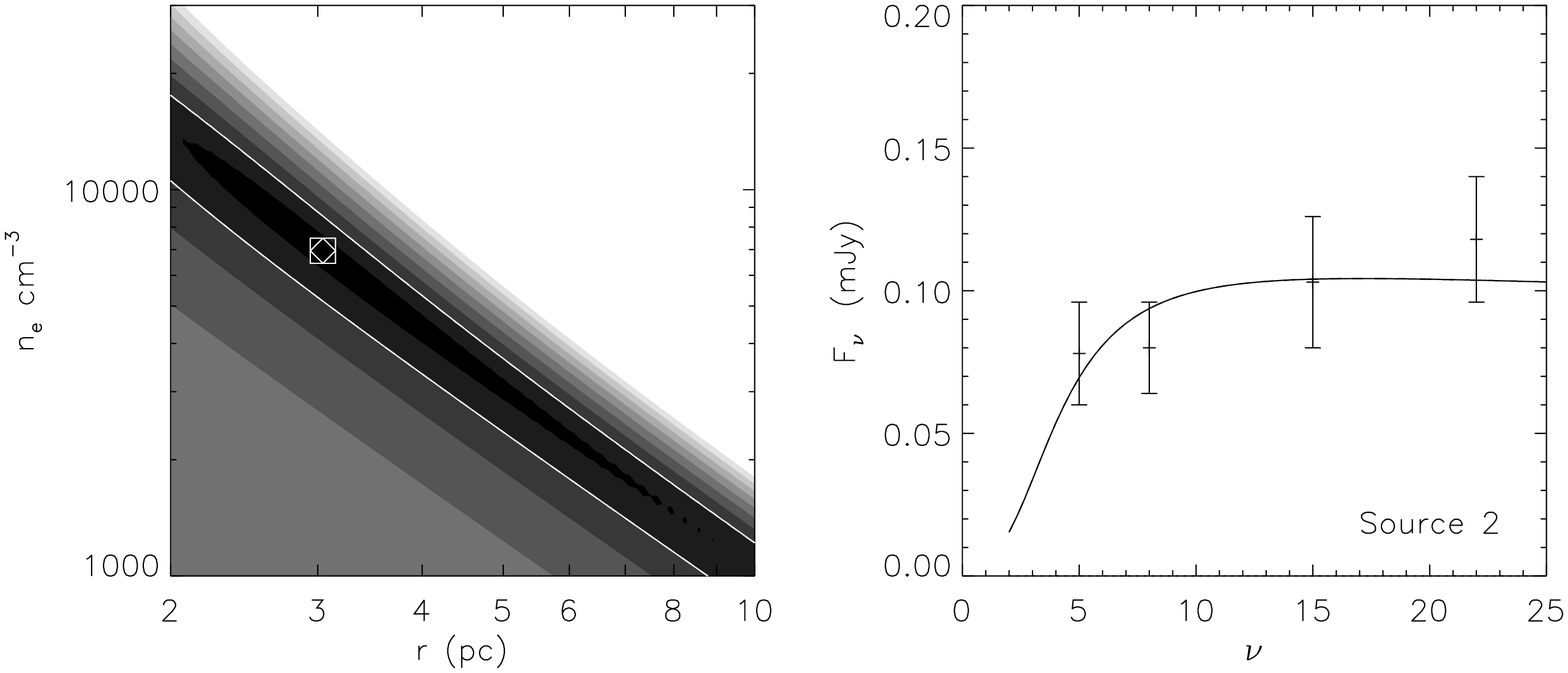}
\plotone{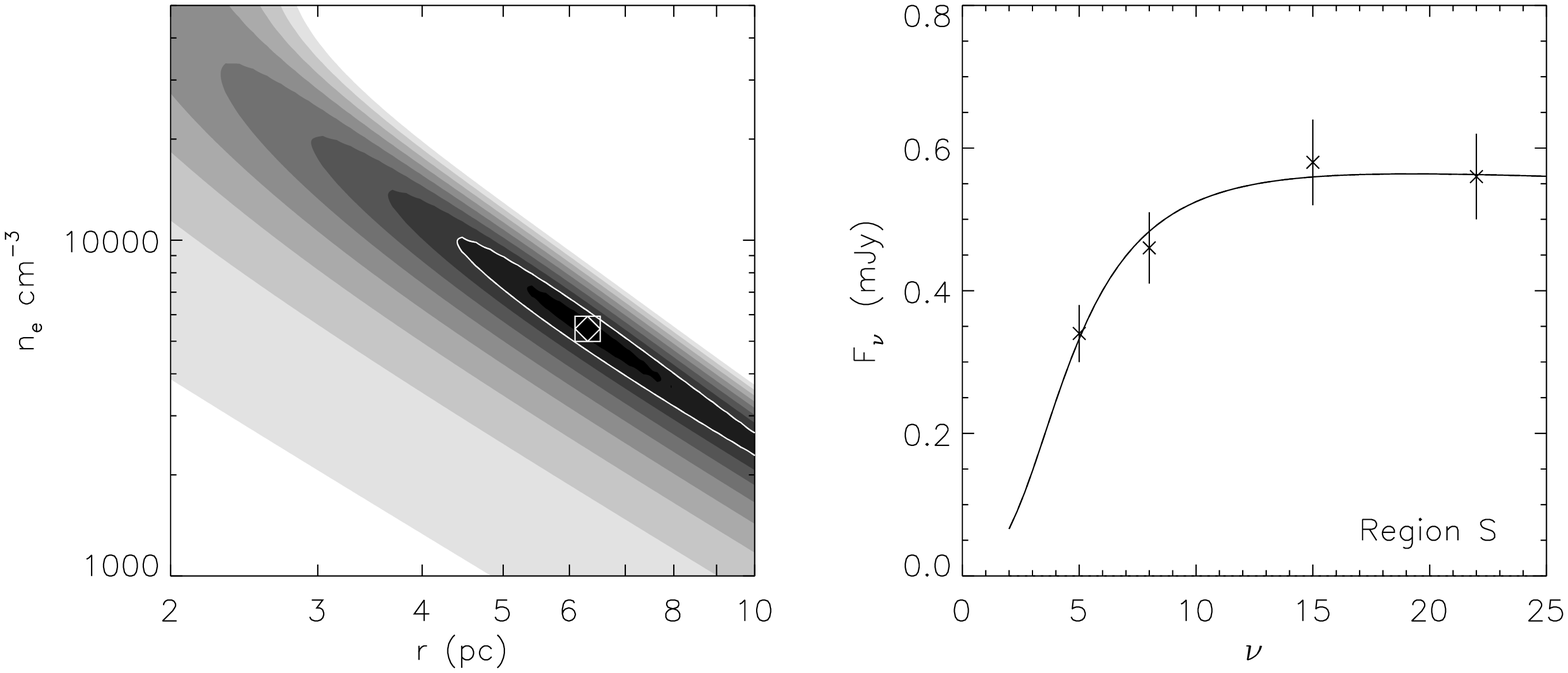}
\caption{Results from model fits to SSC 1, SSC 2, and Region~S (from top to
  bottom, respectively).  (left) Contours of the minima in reduced  
$\chi^2$; a clear degeneracy between radius and density is apparent.
  The diamonds indicate the best-fit purely thermal model, and the
  squares indicate the best-fit model if a non-thermal contribution is
  allowed. (right) The best-fit model SEDs
  over plotted on the observed data points.  Only SSC 1 has a
  marginally improved fit by allowing for a non-thermal contribution.}
\label{plot_S1}
\end{figure*}

The best-fit models with only thermal emission are shown as solid
curves in Fig.~\ref{plot_S1}.  The models also allow for an additional
non-thermal contribution, although this only resulted in a marginally
better fit for for SSC 1 and is shown as a dashed line.  The best-fit
parameters and the physical properties inferred from these models are
reported in Table~\ref{tab:properties}.  However, as can be seen in
Figures~\ref{plot_S1}a-c, it must be stressed that in all cases
presented here, there is a range of parameters that fit the data with
nearly the same goodness-of-fit.  It is important to understand the
degeneracy inherent in fitting thermal radio SEDs of unresolved
sources.  Specifically, at high frequencies where the resolution is
sufficient to separate star-forming regions, sources also tend to be
optically thin or only slightly optically-thick.  In this regime,
radius and density are essentially degenerate.  From the plots of the
reduced $\chi^2$ minima shown in Figures~\ref{plot_S1}a-c, it is clear
that a {\it range} of values for radius and density are consistent
with the data in hand; thus the ``best-fit'' values quoted in
Table~\ref{tab:properties} should not be strictly interpreted.

\begin{deluxetable}{llccccccccccccccc}
\tabletypesize{\footnotesize}
\tablecaption{Properties of the Thermal Radio Sources Inferred from the Spectral Fits\label{tab:properties}}
\tablewidth{0pt}
\tablehead{
& & \colhead{r \tablenotemark{b}} & \colhead{n$_e$ \tablenotemark{b}} 
& \colhead{$Q_{Lyc}$ \tablenotemark{c}}  & \colhead{M$_{stars}$\tablenotemark{d}} \\\\
\colhead{Source} & \colhead{$\chi^2$ \tablenotemark{a}} & \colhead{(pc)} & \colhead{($10^3$ cm$^{-3}$)} 
&  \colhead{(10$^{52}$\,s$^{-1}$)} & \colhead{(10$^6$\,\msun) \tablenotemark{d}} \\ 
}
\startdata
S1       & 0.2 & 5.0  &  4.3  & 3.5  & 0.5  \\
S1$_{nt}$ \tablenotemark{e}   & 0.1 & 4.5  &  4.8  & 3.5  & 0.5\\
S2       & 0.6 & 2.8  &  7.9  & 2.5  & 0.4\\
Region S & 0.3 & 7.9  &  3.2  & 11.7 & 1.7\\
\enddata 
\tablenotetext{a}{It is clear from the reduced $\chi^2$ values that
  these fits are not well-constrained.  See \S~\ref{sec:fits} for
  discussion.}  
\tablenotetext{b}{See Figure~\ref{plot_S1} for range
  of acceptable values.} 
\tablenotetext{c}{{\qlyc\ values as determined
    from the 1.3~cm flux densities, assuming an \hii\ region
    temperature of 20000\,K.} } 
\tablenotetext{d}{Adopting the models
    of SB99 for a burst at age 1Myr \citep{leitherer99}.}
  \tablenotetext{e}{Source SSC 1 had a marginally better fit for models
    that included a small amount of non-thermal emission.}
\end{deluxetable}

The radii of the two thermal sources associated with SSCs 1 and 2
inferred from our fits are $\sim 3-6$\,pc (see Table
\ref{tab:properties}).  These sizes are several times smaller than the
radii of 8-15\,pc deduced by \citet{hunt04} from lower-resolution
observations, but typical of SSCs.  They are also considerably smaller
than those measured by \citet{thompson08} with NICMOS at 1.9\,\micron\
(16--18\,pc).  However, the diffraction limit of \hst\ at this
wavelength of $\sim$0.2\,arcsec (54\,pc) makes direct measurements
with NICMOS potentially unreliable.  Likewise, the radii inferred here
are not directly measured, and therefore these sizes should be
considered estimates.  Moreover, given the youth of these clusters,
they are unlikely to be dynamically relaxed into simple spherical
systems, but rather complex structures that are not resolved by any of
the available data.  

The inferred EMs of $2-4\times10^8$\,pc\,cm$^{-6}$ are about 10 times
higher than that inferred globally.  Hence, the electron densities
$n_e$ of $\sim4-7\times10^{3}$\,cm$^{-3}$ on small spatial scales are
several times greater than those estimated from global radio fluxes,
and roughly 10 times those derived from optical spectra
\citep{izotov99}.  The sensitive high-resolution measurements
presented here are able to probe more deeply into the optically-thick
star-forming region in \sbs, and support the idea that beam dilution
seriously affects low-resolution radio observations of such sources.
Moreover, optical wavelengths may not be sampling the same regions as
these radio observations; long wavelengths are better able to examine
the dusty dense zones in extreme modes of star formation such as those
in \sbs.

The model fits show that the radio emission appears to become
optically thick at frequencies $\lesssim$5-8\,GHz.  Even at 8.5\,GHz,
the best-fit models suggest that \tauff$\sim$0.3-0.5.  Nevertheless,
the highest frequency observations presented here at 22~GHz are
consistent with being optically-thin, and we do not believe that the
flux densities at this frequency have suffered from any significant
self-absorption.  The observed flux densities may, however, suffer
from losses due to other factors. First, dust within the \hii\ region
can absorb ionizing photons, thereby reducing the amount of ionized
gas.  Second, if the ISM is porous (which we will return to in
\S~\ref{sec:clumping}), a significant fraction of the ionizing flux
may escape from the region.

\subsection{Ionizing Luminosities \label{sec:ionizing}}

Thermal radio emission can be used to estimate the production rate of
ionizing photons from the massive star clusters powering these
regions, and the ionizing luminosities can in turn be used to estimate
their stellar content.  

Following \citet{rubin68} and \citet{condon92}, 
\begin{eqnarray}
{{Q_{Lyc}}}&
\geq6.32\times10^{52}~{{\rm s}^{-1}} ~\left(0.926\right)~ 
\Big({{T_e}\over{10^4~{\rm K}}}\Big)^{-0.45}
\Big({{\nu}\over{{\rm Ghz}}}\Big)^{0.1} \nonumber \\
&\times \Big({{L_{thermal}}\over{10^{27}~{\rm erg~s^{-1}~Hz}^{-1}}}\Big).
\end{eqnarray}
The factor 0.926 corrects for the $\sim$8\% helium abundance in the
ionized gas \citep{mezger67}.  This equation assumes the emission is
both thermal and optically thin, as both of these criteria are most
likely to be met at the highest frequency observations, which are less
likely to contain contaminating non-thermal flux and less likely to
suffer from self-absorption.  We use the 1.3\,cm flux densities in
order to determine \qlyc\ for the thermal radio sources in \sbs.  We
have assumed an ionized gas electron temperature of $T_e$=20000\,K
\citep{izotov99}.  Given that an O7.5V star produces an ionizing flux
of \qlyc = $10^{49}$~s$^{-1}$ \citep[hereafter O*, the equivalent to
type O7.5V:][]{leitherer90,vacca94,vacca96}, from Table
\ref{tab:properties} we infer that regions SSCs 1 and 2 contain a
minimum of approximately 3500 and 2500 equivalent O* stars,
respectively.  The entire region S is subject to an ionizing flux from
the equivalent of on the order $\sim$11,700 O* stars.

These \qlyc\ values could very well be underestimates if a significant
fraction of the ionizing flux is absorbed by dust or able to escape
from the \hii\ region, possibly due to clumping in the ISM, which we
believe to be an issue for these objects \citep[see
  \S~\ref{sec:clumping} and ][]{reines08}.  Such an effect has been
observed for Galactic ultracompact \hii\ regions
\citep[e.g.,][]{kurtz99,kim01}, for which it is estimated that
typically $\gtrsim 80$\% of the ionizing flux from the embedded stars
escapes to an outer diffuse halo. We can estimate an upper limit to
this effect for sources SSCs 1 and 2 in \sbs\ by assuming that any flux
able to leak from these regions is still contained in the larger
region S.  This comparison suggests that $\sim 50$\% of the ionizing
flux may be able to escape the immediate vicinity of SSCs 1 and 2, and is in
reasonable agreement with the escape fraction of $\sim 40$\% estimated by
\citet{reines08}.

\subsection{Age of the SSCs\label{sec:age}}

The ages of the star clusters in \sbs\ are not well constrained by our
observations.  However, the existence of thermal radio emission from
only SSCs 1 and 2 suggests that they are younger than the other
four SSCs detected in the optical with {\it HST}.  Several other
observations support the extreme youth of SSCs 1 and 2.  

For example, \citet{vanzi00} observe a very large \brg\ equivalent
width, one of the highest ever observed for an extragalactic object.
While none of the clusters show \lya\ emission, SSCs 4 and 5 are
considerably brighter than SSCs 1 and 2 in the far-ultraviolet
continuum \citep{kunth03}.  This is consistent with SSCs 1 and 2
having younger ages and thus suffering from more extinction in the
natal cloud.  More recently, \citet{reines08} find that SSCs 1 and 2
have ages $\lesssim 3$~Myr based on fitting model SEDs to optical
photometry and measuring H$\alpha$ equivalent widths, which is
consistent with the earlier work of \citet{thuan97}.  We find that the
Pa$\alpha$ equivalent width is not a reliable tracer for age due to
the near-IR excess and free-free emission that can significantly
contaminate the continuum \citep{reines08}.  The upper limit of
$\sim$3~Myr arises because hydrogen emission-line widths and stellar
colors are insensitive to differences in ages less than this, while
the ionizing flux is roughly constant before the most massive stars
have started to die.  Moreover, the stellar synthesis models are not
well-calibrated at these extremely young ages.  This upper limit of
3~Myr agrees with previous radio studies \citep[e.g.][]{kj99} that
conclude that the SSCs which are detectable as optically-thick thermal
radio sources are $\lesssim 1$~Myr (based on statistics).  SSCs 3-6,
which are not detected in the radio, have ages ranging from $\sim
7-15$~Myr \citep{reines08}.

The seemingly insignificant difference in age of a few Myr for an SSC
could actually be important; $\sim3-4$\,Myr is a very ``special''
epoch in the life of a young starburst.  SSCs and their environment
undergo a tremendous amount of development between 1~Myr and 6~Myr,
making this age range important to target for studying the details of
their early evolution.  First, this age is just after the onset of
significant Wolf-Rayet (W-R) populations at $\sim3$\,Myr \citep{leitherer99};
the W-R component augments by a factor of ten or so the amplitude of
the cluster stellar wind \citep{leitherer92}.  Second, at
$\sim$3.5\,Myr, Type II SNe begin to explode, further disrupting the
ISM through shocks, and increasing still more the amount of
thermalized gas streaming out of the confines of the star cluster
\citep[e.g.][]{chevalier85}.  This is the age when the ISM is
subjected to an abrupt change in mechanical energy input because of
the SN contribution, and to a lesser extent from the W-R winds.
The presence of a small, but measurable, W-R population in SSC 3
\citep{papaderos06,izotov06}, but not in SSCs 1 and 2, is a confirmation
of the younger age of these latter clusters.

During this short period of only a few million years, the clusters
will transform from being optically obscured and extremely luminous in
the infrared, to completely optically visible with little to no
extinction.  Very young sources are embedded in dense cocoons of gas
and dust and show rising radio spectra which gradually flatten and
become fainter as they become increasingly optically thin with time
\citep{cannon04,johnson04,hirashita06}.  The clusters at ages of
$\lesssim$4\,Myr but older than $\gtrsim$3\,Myr will develop a 
wind component in their SEDs, unlike younger clusters without W-R
stars and SNe.  At lower frequencies, the spectra should begin to
steepen around $\sim$3.5\,Myr because of the non-thermal emission from
Type II SNe, and finally become predominantly non-thermal as the
thermal \hii\ emission dies out at $\sim$10\,Myr.  Thus, with
sufficient sensitivity, the radio spectral energy distribution can
also serve as a diagnostic of a cluster's evolutionary stage.  As
discussed above, the maximum age that SSCs 1 and 2 could have is
$\sim3$\,Myr, and they could be as young as $\sim 1$~Myr.  By
contrast, SSCs 4 and 5 are $\sim 12-15$~Myr \citep{reines08}, and
fully optically visible with no detected radio or thermal infrared
emission.

Given the non-thermal emission detected on large scales by
\citet{hunt04}, it is clear that previous episodes of star formation
have taken place in \sbs.  However, at the present time we are
not able to precisely associate the non-thermal emission with a
specific cluster or clusters.

\subsection{Stellar Content and Star Formation Rates}

The stellar content of the radio \hii\ regions can be estimated from
their Lyman continuum luminosities (neglecting any leakage or dust
absorption) using the Starburst99 models of \citet{leitherer99}.  For
a cluster $\lesssim$3~Myr old, formed instantaneously with a Salpeter
IMF, 100~$M_\odot$ upper cutoff, 1~$M_\odot$ lower cutoff (note that
reducing the lower mass cutoff to 0.1~$M_\odot$ increases the cluster
mass by a factor of $\sim 2.5$), and 5\% solar metallicity, a $10^6
M_\odot$ cluster has \qlyc$\simeq 7.4 \times 10^{52}$~s$^{-1}$.
Assuming that \qlyc\ scales directly with the cluster mass, we find
that the thermal regions around SSCs 1 and 2 are powered by stellar
clusters with masses $\sim
5\times10^5$\,\msun\ (Table~\ref{tab:properties}).  Since the ionizing
luminosity is roughly constant from 1-3~Myr, this result is
independent of age, as long as the age of the clusters is
$\lesssim$3~Myr.  These masses are roughly a factor of two lower than
those found by \citet{reines08}, because of the underluminosity of the
radio emission relative to the optical SEDs (see
$\S$\ref{sec:clumping}).

In these young massive clusters, the number density of just the O*
stars is high: $\sim$3000 such stars in a sphere of 4\,pc radius gives
$\sim$11\,pc$^{-3}$.  These densities are such that the extent of a
typical stellar wind would significantly overlap with that of its
neighbor in less than 1\,Myr \citep[e.g.,][]{weaver77}.  The effect of
these ``colliding winds'' should be strong X-ray emission
\citep{pittard97,zhekov93}, over and above the diffuse X-ray emission
expected from the adiabatic interaction of the stellar wind
\citep{chevalier85,canto00}.  Compact X-ray emission is indeed
observed in \sbs\ \citep{thuan04}, although the spatial resolution of
those observations is insufficient to associate it only with SSCs 1
and 2.

At an age of $\lesssim$3\,Myr, the individual clusters SSCs 1 and 2 would
together provide a bolometric luminosity of \lbol$\sim1.5-1.9\times10^9$\lsun,
according to the models of Starburst99 for an instantaneous burst with
a metallicity of 5\% solar.  This is in reasonable agreement with the
infrared luminosity of \sbs\ estimated from the dust emission,
$1.4-1.5\times10^9$\lsun\ \citep{hunt05a,engelbracht05}.  The
additional ionizing luminosity in the entirety of region S would yield 
\lbol$\sim3\times10^9$\lsun.  This excess at larger scales could mean that
the dust is intercepting only about 50\% of the ionizing radiation
globally.  Consequently, the bolometric luminosity inferred
from infrared observations could be underestimated.

The instantaneous star formation rate (SFR) can be estimated using the
$Q_{Lyc}$ luminosity.  Following \citet{kennicutt98}:
\begin{equation}
{{{\rm SFR} (M_\odot~{\rm year^{-1}})}}
= 1.08 \times10^{-53}~Q_{Lyc}{({\rm s}^{-1})}.
\end{equation}
The inferred $Q_{Lyc}$ value from the 1.3~cm emission yields a
SFR$\simeq 1.3 M_\odot$yr$^{-1}$ for region~S (270\,pc in diameter),
which is quite substantial.  
Alternatively, with an upper limit to the age ($\lesssim$3\,Myr) and
an estimate of the stellar mass of the SSCs
($\sim2\times5\times10^5$\,\msun, corrected by a factor of 2.5 to
extend the IMF down to $0.1M_\odot$ and dividing by a factor of 0.6 to
roughly correct for the fraction of escaping radiation), we can simply
determine an average SFR, and find $\sim$1.4\,\msun\,yr$^{-1}$.  By
comparison, using \ha\ emission, \citet{thuan97} find a SFR for the
entire galaxy of $\sim 0.4 M_\odot$~yr$^{-1}$, already among the
highest SFRs observed in BCDs \citep{fanelli88}.  This discrepancy may
suggest that a substantial portion of the current star formation that
is traced by thermal radio emission in \sbs\ is hidden from view in
optical recombination lines \citep{hunt01}.  However, there is at
least a $\sim 30$\% dispersion between SFR calibrations from different
authors \citep[e.g.][and references therein]{kennicutt98}, which could
partially account for the observed discrepancy.  This is similar to
the difference in star formation derived from optical and radio
observations for the BCD Haro~3 \citep{johnson_etal04}.  Clearly the
nascent star formation in region~S is a major event for this dwarf
galaxy, and serves to highlight the impact a small region of star
formation can have on a small dwarf galaxy.

The relevance of such an impact can be perhaps better appreciated by
considering that this relatively high (for a dwarf galaxy) star
formation rate takes place in a very small region.  Considering a
volume with radius $R\simeq$135\,pc (as for Region~S), we would derive
a star-formation per unit area of
$\sim$23\,\msun\,yr$^{-1}$\,kpc$^{-2}$.  This is comparable to,
although lower than, the starburst intensity limit of
45\,\msun\,yr$^{-1}$\,kpc$^{-2}$ derived by \citet{meurer97} from
observations of starbursts at various wavelengths.  Star formation in
the SSCs in \sbs\ is occurring very close to the maximum intensity
observed in the universe, and, unlike larger systems, this mode of
star formation dominates the energy output of the entire galaxy.

\section{The Starburst in \sbs\label{sec:general}}

In what follows, we discuss the consequences of our results in the
context of the galaxy itself, and their impact on our knowledge of how
stars form in low-metallicity environments.

\subsection{The Porous Circum-cluster Medium \label{sec:clumping} }

Several different results suggest that the ISM surrounding the natal
clusters in \sbs\ is porous and clumpy.  First, the spectral fit of
Region~S suggests that there is ionizing radiation beyond the strict
boundaries of the SSCs 1 and 2.  The volume emission measure \nesqv\
and the number of ionizing photons $Q_{Lyc}$ (see
$\S$\ref{sec:ionizing}) of region S are nearly twice as large as the
sum of SSCs 1 and 2 (Table \ref{tab:properties}).  This implies that
there are ionized zones outside of the strict confines of the SSCs
that contribute to the radio emission.  The extended emission could be
caused by a low level of distributed star formation.  Alternatively,
it could be caused by ionizing radiation that has escaped from the
compact \hii\ regions via a porous ISM.  This effect due to ionizing
radiation leakage has also been observed for ultra compact \hii\
regions in the Milky Way \citep{kurtz99, kim01}.

In previous work, \citet{thuan05} inferred a clumpy ISM in \sbs\ based
on their H$_2$ absorption-line observations.  More recently,
\citet{reines08} also reveal evidence for a porous and clumpy ISM
surrounding the young SSCs 1 and 2: the measured ionizing luminosities
from \ha, \pa, {\it and} optically-thin free-free radio emission are
lower than expected compared to the optical SEDs.  Reines et al. infer
that $\sim 40\%$ of ionizing photons from the stellar continuum are
leaking out of the immediate vicinity of the clusters before
intercepting hydrogen to ionize.  Thus, a significant fraction of the
ionizing flux is escaping without contributing to the measured ionized
gas emission.

Finally, despite the substantial extinction derived from mid-IR
observations toward this region \citep{thuan99,
plante02,hunt05a,houck04}, some UV continuum light is still able to
escape \citep{kunth03}.  Moreover, optically-derived extinctions are
quite low \citep{izotov97,reines08}.  This would naturally follow from
inhomogeneities in the surrounding dust that would simultaneously
allow for large extinctions derived from mid-IR emission originating
from dense dust clumps (as opposed to a uniform screen) and low
extinction regions from relatively transparent lines-of-sight into the
birth cloud \citep[see][for a full analysis and discussion]{reines08}.

Given the impact that clumping of the ISM can have on the resulting
observed spectral energy distribution, as seen in this case for \sbs\,
it is important to keep the potential effect of clumpiness and
inhomogeneities in mind when interpreting similar data.

\subsection{Toward Reconciling the Inferred Extinction \label{sec:other_obs}}

The nature of the observed and predicted line ratios in \sbs\ has been
a source of discussion in the literature
\citep[e.g.][]{hunt01,hunt04,thompson06, reines08, thompson08}.  In
particular, an apparent discrepancy between the thermal radio emission
and infrared line fluxes was first discussed by \citet{hunt04} using
global fluxes, and by \citet{thompson06} using a preliminary version
of the radio observations now published here.  The intrinsic thermal
flux in \citet{hunt04}, inferred from fitting the radio SED, is about
50\% lower than we find with these new observations, and thus appeared 
to be under-luminous with respect to previous near-infrared (NIR) 
recombination line measurements (e.g., \bra\ and \brg).  The unpublished 
radio flux densities used by \citet{thompson06} were derived from a smaller
aperture than used in this paper, consequently resulting in lower
flux densities.  Using the updated values, we find that the observed
and predicted \pa\ values ($1.77\times 10^{-14}$ and $4.05\times
10^{-14}$ erg~s$^{-1}$cm$^{-2}$, respectively) are in good agreement
for a native extinction of A$_{{\rm Pa}\alpha} = 0.9$, which is
significantly lower than the A$_{{\rm Pa}\alpha} = 1.64$ adopted by
Thompson et al.  \citep[see][for further discussion]{reines08}. Thus,
we conclude that observed emission line fluxes and line ratios are
fully consistent with the flux densities obtained from the new radio
observations presented here.

\subsection{Stellar Winds? \label{sec:winds}}

\citet{thuanizotov97} found evidence for P~Cygni profiles in \sbs,
integrated over the 2$\times$2 arcsec$^2$ aperture of the Goddard
High-Resolution Spectrograph aboard the \hst.  This aperture is
sufficiently large to encompass all 6 SSCs in \sbs, so it was
difficult to pinpoint their origin.  The most likely scenario seemed
to be that they arise from massive stellar winds in the older clusters
to the northwest \citep{thuanizotov97}, rather than the younger ones
to the southeast. This conclusion is supported by
later observations of wider \ha\ profiles in SSCs 4 and 5 than in SSCs
1 and 2 \citep{izotov06}. 

More recently,
\citet{thompson06} interpreted the radio flux densities of SSCs 1 and
2 as resulting from a dense stellar wind causing self-absorption at all 
observed radio wavelengths.  If there were such a wind powering the radio
emission in \sbs, we would first expect no spectral turnover at the
highest frequencies observed, in contrast with the spectral energy
distribution of region~S.  We would also expect significant NIR
recombination line excesses relative to ``normal'' \hii-region
emission.  Infrared recombination lines are expected to show the
effect of dense stellar winds much more than the radio, because the
radio signature of such winds is at least two orders of magnitude
weaker than the effect on NIR recombination line opacity
\citep[e.g.][]{smith87,simon83}.  Such an effect is observed for the
massive O stars in the Galactic Center Arches Cluster
\citep{nagata95,lang01}.  Relative to the ratio in \hii\ regions, the
IR-line-to-radio-continuum ratio for those stars is $\gtrsim$250; in
other words, {\it the IR line emission generated from ionized winds
  around massive stars is roughly 250 times stronger relative to the
  radio continuum than it would be in an \hii\ region}.  This is
entirely consistent with stellar wind theory
\citep[e.g.,][]{krolik81,simon83,smith87}.  However, NIR line excesses
are not observed in \sbs\footnote{\citet{hunt04} found such an excess,
  but their thermal radio flux is half of what we deduce from the new
  high-resolution observations presented here.}.  Because of the
consistency with our new radio observations and the NIR recombination
lines \citep[e.g.,][]{reines08}, we conclude that SSCs 1 and 2 do not
show evidence for massive stellar winds.  The lack of observed winds
is consistent with both their young age of $\lesssim$3\,Myr (no W-R
population) and low metallicity \citep[mass loss goes roughly as
$Z^{-0.5}$, see][]{abbott82,leitherer92,crowther02}.  
Because the combined effects of young age and low metallicity
conspire to reduce wind power, a lack of winds alone does not 
constrain the ages of SSCs 1 and 2 to be young.
However, winds have been observed in other clusters in \sbs\ (see above), 
which suggests
that the difference could be the older age of those SSCs.

\subsection{Dearth of Non-Thermal Emission at High Resolution \label{sec:nt}}

The data presented here are fully consistent with purely thermal
models.  The results of non-thermal models only give a marginally
better fit for source SSC 1 than the purely thermal models, and at the
cost of additional free parameters.  Nevertheless, if non-thermal
emission is included in model fits, S1 is consistent with having a
non-thermal fraction at 5~GHz of $\sim 5$\% with a non-thermal slope
of $-0.8$.  The best fits for SSC 2 and Region~S are purely thermal.
In order to include a sufficiently weak non-thermal (synchrotron)
emission in the model fits, these data could at most support roughly a
single SN observed soon after explosion, or alternatively a larger
number of SNe at somewhat later epochs.  

None of the SSCs in \sbs\ have detectable non-thermal radio emission
in the data presented in this paper.  At best, SSCs 5 and 6 have
marginal radio detections at 1.3~cm.  SSCs 4, 5, and 6 all appear to
be sufficiently old that they are likely to contain a number of SNR.
However, based on a comparison with the predictions of
\citet{hirashita06}, the expected non-thermal flux originating from
these clusters is slightly below the sensitivity limits of these data, 
which are relatively high-frequency and not optimized for the detection 
of non-thermal sources.

{\it If} the observations were to definitively detect the presence of
such non-thermal emission from these natal clusters (assuming it were
not due to a projection or confusion effects), it would serve as a
strong lower limit on the age of the star cluster.  Stellar
evolutionary models predict that no SNe could have exploded before
$\sim$3.5\,Myr \citep{leitherer99}.  This age would be in
contradiction to the results from \citet{kj99}, which suggest (based
on statistics) that the lifetime of the ``ultradense \hii\ regions''
is $\lesssim 1$~Myr.  In addition, upper limits on the ages of SSCs 1
and 2 of $\lesssim 3$~Myr were determined by \citet{reines08} using
optical and infrared observations; these results provide additional
support for the young ages of these clusters, and thus the expected
lack of non-thermal emission.  

\begin{figure}[t]
\epsscale{1}
\plotone{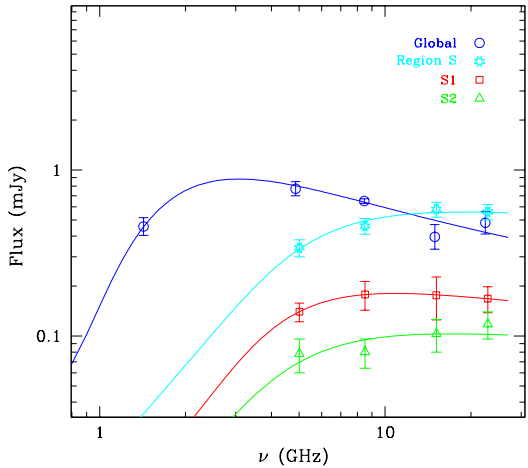}
\caption{A comparison between the flux densities observed at
  high-resolution (this paper) and the flux densities observed by
  \citet{hunt04}, which are sensitive to larger spatial scales.  The
  different symbols (open circle, star, square, and triangle)
  correspond to the global (low-resolution) data, Region~S, SSC 1, and
  SSC 2, respectively.  The best-fit models are also shown: the shape
  of the global spectrum is indicative of a strong non-thermal
  component.
\label{plot_hunt}
}
\end{figure}

These results may appear to be in contradiction with previous results
from \citet{hunt04}, in which a non-thermal component was detected.
However, \citet{hunt04} obtained lower resolution observations of
\sbs\ that are sensitive to larger-scale spatial structure than the
observations presented here (by virtue of interferometers acting as
spatial filters).  The flux densities derived from the lower
resolution observations are over-plotted on the high-resolution
observations in Figure~\ref{plot_hunt}.  The Hunt et al. data clearly
indicate the presence of non-thermal emission in \sbs, as manifest in
the negative slope of the spectral energy distribution.  A comparison
between the two data sets demonstrates that the observations presented
here are resolving out a substantial fraction of the longer wavelength
emission at 6~cm \citep[roughly 44\% of that observed by][]{hunt04},
while recovering all of the emission at shorter wavelengths.  The
unresolved 20~cm source in the lower resolution observations is also
offset to the northwest of SSCs 1 and 2 (see Figure~\ref{LcontK}), but
there is no evidence in those observations for a peak in non-thermal
emission at the location of the thermal radio sources.  However, the
center of the diffuse non-thermal emission could possibly coincide
with a faint optical source, visible in the ACS image toward SSC 3.

We ascribe this difference to the diffuse nature of synchrotron halos
around massive star clusters \citep[e.g.,][]{cannon05,hunt05b}.  The
dearth of non-thermal emission in these high-resolution observations
suggests that the non-thermal emission is diffuse in nature, and not
obviously associated with the compact thermal radio sources we detect.
If the synchrotron halo of I\,Zw\,18 were to be placed at the distance
of \sbs, it would be resolved out by our high-frequency observations
(LAS$\sim$2\arcsec).  Such halos are dominant at low frequencies and
have low surface brightness, which makes them difficult to detect with
the observations presented here.  Hence, we conclude that our
observations are not inconsistent with previous ones.  Moreover, the
presence of non-thermal emission in close vicinity to the compact
thermal radio sources also suggests that significant star formation
has taken place preceding the birth of sources SSCs 1 and 2.  

\subsection{The Importance of Metallicity in Massive Star Cluster Evolution}

Natal super star clusters presumably emerge from their birth cocoons
via a combination of expansion of the high pressure regions, radiative
dissociation of molecules and dust, and stellar winds.  The interplay
of these processes will affect the timescales in the evolution of
natal clusters, and in particular how rapidly a cluster emerges from its
birth material.  Each of these processes has a dependence on
metallicity, and thus the low metallicity of \sbs\ could
potentially have a strong combined effect on the early evolution of
the natal clusters that it hosts.

First, the pressure of the ionized gas within the cluster may be
affected by metallicity.  Metallicity has a fundamental role in the
rapidity of radiative cooling of the ISM within and outside the
cluster \citep{mccray87,silich04,tenorio05a}, and therefore the
pressure of the region will be affected.  The average pressure of the
ionized gas in the clusters implied by the densities we infer from the
spectral fits is quite high, $\sim1.4\times10^{-8}$\,dyne\,cm$^{-2}$.
Given the ionized gas electron temperature of $T_e$=20000\,K
\citep{izotov99} is roughly twice the ``typical'' value expected for
\hii\ regions, it follows that the pressure would be correspondingly
twice as high as that found in a typical \hii\ region.  A comparison
with other galaxies hosting natal clusters suggests that the inferred
pressure for the SSCs 1 and 2 in \sbs\ is similar to in other
SSCs in dwarf galaxies \citep[e.g., \hen, IC\,4662, NGC\,5253,
II\,Zw\,40][]{jk03,johnson03,turner00,beck02}.  However, given the
uncertainties in the model fits to the spectral energy distributions,
the inferred pressures are, at best, accurate to within a factor of
two. More precise measurements of the pressure in natal SSCs from a 
range of environments will help to address this issue. 

The metallicity can also affect the abundance of molecules and dust in
the natal cocoon.  First, at low metallicities, the relative amount of
molecules and dust is expected to be lower than in higher metallicity
counterparts.  Moreover, the dust itself has a critical role in the
formation and survival of molecules 
\citep[e.g.][]{hollenbach71,salpeter77,savage77,hollenbach97},
thus there is an intimate connection between metallicity,
dust, and molecules.  In addition, lower metallicity systems will also
have harder radiation fields and create a more hostile environment for
molecules and dust \citep[e.g.][]{madden06,lebouteiller07}.  These
effects would tend to shorten the timescale for a natal SSC to emerge
from its birth material.  SSCs 1 and 2 in \sbs\
are already associated with some optical emission, consistent with
this scenario, and it is possible that the low metallicity environment
has accelerated the emergence process.  However, the partial emergence
of these clusters could also simply reflect their nominal evolutionary
state, and we are not able to discriminate between these possibilities
with existing data.

Another mechanism by which metallicity may affect the early evolution
of massive clusters is via their stellar winds.  It is well known that
the effect of massive stellar winds decreases with metallicity
\citep{abbott82,leitherer92,vink01,kudritzki00,crowther02}, primarily
due to reduced line blanketing.  Even before the onset of the W-R phase
and the first supernova in a cluster, the mechanical energy input
from the massive star winds alone is more than an order of magnitude
weaker for systems with metallicity as low as \sbs\ compared to
solar metallicity counterparts \citep{leitherer99}.  In contrast to
the effects discussed above, the reduced mechanical energy input for a
low metallicity cluster would tend to lengthen the embedded phase --
radiation pressure would not be able to clear away the surrounding
material as effectively.  However, at the present time, it is unclear
which (if any) of these effects will have a dominant role in the early
evolution of massive star clusters with low metallicity.

\section{Summary \label{sec:summary}}

This paper examines the radio continuum properties of current
star-forming regions in the extremely low-metallicity galaxy
\sbs\ using high-resolution VLA observations.  Two intense
star-forming regions are detected as luminous thermal radio sources
that appear to be extremely young SSCs.  The total ionizing flux of
the southern star-forming region in the galaxy suggests the presence
of $\sim 12,000$ ``equivalent'' O-type stars, and the inferred
instantaneous star formation rate for the radio-detected natal star
clusters {\it alone} is $\sim 1.3$~M$_\odot$yr$^{-1}$ or $\sim
23$~M$_\odot$yr$^{-1}$kpc$^{-2}$, which is close to the starburst
intensity limit of $45$~M$_\odot$yr$^{-1}$kpc$^{-2}$.  This star
formation rate derived from thermal radio emission also suggests that
previous optical recombination line studies are not detecting a
significant fraction of the current star formation in SBS~0335-052.
The observations presented here also suggest that up to $\sim 50$\% of
the ionizing flux could be leaking out of the compact \hii\ regions;
this in is agreement with previous work that suggests the interstellar
medium surrounding the natal clusters in \sbs\ is porous and clumpy.
Model fits to the radio spectral energy distribution indicate that the
{\it global mean} density in the youngest SSCs is n$_e \gtrsim
10^3-10^4$~cm$^{-3}$.

A comparison between these radio data and those of \citet{hunt04}
suggest that the observations presented here have resolved out a
significant contribution from non-thermal emission over spatial scales
not detected in these high-resolution observations. The presence of
this diffuse non-thermal emission may indicate a synchrotron halo in
the southern region of \sbs, likely associated with a previous
episode of star formation in proximity to the natal clusters detected
here that has not yet been identified.  Such a previous generation of
massive stars will have a rapid impact on the chemical enrichment in
this very low-metallicity galaxy.

A fundamental question is how SSCs form and evolve at low
metallicities.  While we cannot presume to answer this with these
observations alone, it is clear that \sbs\ is undergoing an
extreme burst of star formation, the results of which are likely to
significantly impact the galaxy's future evolution.  The low
metallicity in this galaxy may affect the evolution of the natal
clusters in a variety of ways, including inefficient cooling, hard
radiation fields, and weak radiation pressure from line blanketing.
However, a larger sample of low-metallicity natal clusters with
panchromatic observations will be required to address these issues.
These observations do, however, conclusively demonstrate that
massive star clusters are indeed able to form in low-metallicity
environments, which is an important step toward our understanding of 
globular cluster formation in the early universe.

\acknowledgements We are grateful to an anonymous referee for 
insightful suggestions, and to Trinh Thuan, Yuri Izotov, and
Rodger Thompson for many useful conversations.  We thank St\'ephanie
Plante for her input during the early stages of this project.  KEJ
gratefully acknowledges support for this paper provided by NSF through
CAREER award 0548103 and the David and Lucile Packard Foundation
through a Packard Fellowship.

\end{document}